\documentclass[preprint,preprintnumbers,amsmath,amssymb,prb]{revtex4}

\usepackage{graphicx}
\usepackage{dcolumn}
\usepackage{bm}
\usepackage{amssymb}
\usepackage{latexsym}

\begin{document}
\title{Simple Model of Membrane Proteins Including Solvent}
\author{D. L. Pagan, A. Shiryayev, T. P. Connor, and J. D. Gunton
      \\    \emph{\small{Department of Physics, Lehigh University, Bethlehem, P.A.
       18015}}}

\begin{abstract}
    \noindent We report a numerical simulation for the phase diagram of a simple
    two dimensional
    model, similar to one proposed by Noro and
    Frenkel [J. Chem. Phys.
     \textbf{114}, 2477 (2001)] for membrane proteins, but one that includes
    the role of the solvent. We first use Gibbs ensemble Monte
    Caro simulations to determine the phase behavior of particles
    interacting via a square-well potential in two dimensions for
    various values of the interaction range. A phenomenological
    model for the solute-solvent interactions is then studied to
    understand how the fluid-fluid coexistence curve is modified
    by solute-solvent interactions. It is shown that such a model
    can yield systems with liquid-liquid phase separation curves
    that have both upper and lower critical points, as well as closed
    loop phase diagrams, as is the case with the corresponding
    three dimensional model.
\end{abstract}

\maketitle

\section{Introduction}

    Although there exists no theoretical framework concerning how
    to optimally grow protein crystals, experimentalists have
    produced a large number of globular protein crystals suitable for x-ray
    diffraction. This is evidenced by the large number of crystal
    structures available in the protein databank~\cite{kn:databank2}.
    The situation, however, is rather
    different for the case of membrane proteins, a class of important
    proteins which are characterized as being attached to a lipid membrane\cite{kn:singer}.
    These proteins have a number of important functions, including
    the catalysis of specific transport of metabolites and ions
    across membrane barriers, the conversion of sunlight energy
    into chemical and electrical energy, and the reception and
    transduction of signals (such as neurotransmitters or
    hormones) across the cell membrane\cite{kn:michel}.
    By comparison, far fewer
    membrane protein structures are known than for globular
    proteins, due to the difficulty in crystallizing membrane
    proteins.  Membrane proteins are  embedded in the
    electrically insulating lipid bilayers of biological
    membranes.  The parts of the protein surface which are in
    contact with the alkane chains of the lipids are highly
    hydrophobic, whereas their surfaces that are exposed to the
    aqueous media on each side of the membrane are hydrophilic.
    This amphiphilic nature of membrane proteins makes it
    difficult to crystallize them, since there are large
    hydrophobic and hydrophilic surface areas.  As a consequence
    they are not soluble in aqueous solutions (unlike globular
    proteins), so that detergents are commonly used to isolate and
    purify them.  The detergents are themselves amphiphilic
    molecules which form micelles at sufficiently large
    concentrations.  The detergent micelles incorporate the
    proteins in the solubilization process.  The membrane protein
    in the detergent micelle is then the initial material used for
    purification and crystallization.   It has been
    shown that membrane proteins can often be more easily crystallized
    if they are forced to interact in a quasi-two dimensional
    space\cite{kn:michel,kn:rigaud}. Many approaches to crystallizing membrane proteins
    utilize this procedure.  Thus it is of interest to know
    the general features of the phase behavior of
    quasi-two dimensional proteins and to compare these with those
    for globular proteins. This is obviously a challenging problem
    due to the additional role of the detergent and solvent in
    mediating the interactions between membrane proteins.

    The phase behavior of globular proteins has often been modelled using ideas from
    a similar problem in colloidal science.
    It has been
    shown\cite{kn:gast,kn:other,kn:hagen,kn:rosenbaum,kn:tenwolde,kn:oxtoby,kn:pagan_2004,
    kn:pagan_2005,kn:shiryayev_2004}
    that to a reasonable first approximation
    these proteins can be modelled using a short-range attractive
    interaction and a hard-core repulsion.  Indirect evidence suggests that
    membrane proteins also interact via a short-range interaction as well.
    As a consequence, a generic phase
    diagram for membrane proteins has been proposed\cite{kn:noro} based on a
    short-range interaction model.  In addition, a similar
    approach\cite{kn:annexinFrenkel} has
    been used to successfully model the phase behavior of the
    protein Annexin V.  However, an important effect that has been
    largely ignored until recently is the role of the
    solute-solvent interactions on such phase diagrams. For three-dimensional
    globular proteins, crystals are grown
    using a combination of a variety of precipitating agents,
    including salts and polyethylene glycol. These agents (along
    with other additives to control such parameters as the pH), are used to fine-tune
     the interactions
    between proteins and control the range and strength of interactions
    between protein particles. Precipitating agents are also important
    in crystallizing membrane proteins.  As a first step toward understanding the
    role of solute-solvent interactions in protein crystallization, a model was proposed recently
    \cite{kn:shiryayev_2005} for globular proteins interacting via
 a short-range  potential interacting in three
    dimensions. In that study, it was shown that a variety of
    phase diagrams could be obtained for different choices of the solute-solvent
    parameters. In this paper
    we extend that study to model particles interacting in a
    quasi two-dimensional plane in the presence of a solvent.  We
    first
    consider the particular case of particles interacting via a square-well potential,
    in the absence
    of solvent,
    determining
    the fluid-fluid coexistence curves for several different interaction
    ranges, using standard Monte Carlo methods.  We then obtain
    the phase diagram for such a system taking into account
    solute-solvent interactions and show that there are several
    possible types of phase diagrams, depending on the choice of
    interaction parameters.  Namely, the model can have an upper
    critical point, a lower critical point, or closed loop phase
    diagrams, depending on the solute-solvent interactions. Our
    work is an extension of the original work of Noro and Frenkel
    \cite{kn:noro} in that we include these solute-solvent
    interactions (in a very simple way).

\section{Computational and Theoretical Details}

    We use a square-well potential to model the attractive
    interactions between protein particles interacting in a 2D
    plane. The square-well potential is given by

   \begin{equation}
        {V(r)} = \left\{ \begin{array}{ccc}
                        \infty, &\mbox{$r < \sigma$}  \\
                        -\epsilon, &\mbox{$\sigma \le r <
                        \lambda\sigma$}\\
                        0, &\mbox{$r \ge \lambda\sigma$},
                   \end{array} \right.
                   \label{square_well}
    \end{equation}

    \noindent where the particle diameter, $\sigma$, and the
    well-depth, $\epsilon$, set the length and energy
    scale, respectively, and $\lambda$ is the interaction range. We attempt to obtain the fluid-fluid
    coexistence curve of particles interacting in a 2D plane via eq. (\ref{square_well}) for the ranges $1.25 \leq \lambda \leq 2.0$.
    To calculate the fluid-fluid coexistence curves, we employ
    the Gibbs ensemble Monte Carlo method\cite{kn:gibbs}. This method is well
    suited to studying such systems away from the critical point
    and is a standard technique.

    As noted in the introduction, membrane proteins are often
    crystallized in two dimensions (2D)  where, as in three dimensions  (3D), various solvents are
    used to promote crystallization. These precipitating agents thus
    play an important part in the crystallization process. As noted in the introduction,  a
    simple phenomenological model for the role of
    solvent has been proposed~\cite{kn:shiryayev_2005}. The multi-component system
    (i.e., proteins plus
    solvent) is modelled as a binary system in which the solute
    particle is much bigger than the solvent. It is also assumed
    that the effect of the other components are subsumed in the
    effective solute-solvent interaction.

    The total energy of a protein system interacting via a
    short-range potential can be written as

    \begin{equation}
        U_{0}(\vec{r}^N) = \frac{1}{2}\sum_{i\neq j} U_0(|\vec{r}_i -
        \vec{r}_j|),
    \end{equation}

    \noindent where the subscript '0' denotes the potential of the protein
    in the
    absence of solvent. The model that incorporates the
    effect of solvent has a total potential

    \begin{equation}
        U = U_0 + \sum (\epsilon_w - k_BT\Delta s_w)
        n_w^{(i)}(\vec{r}^N),
    \end{equation}

    \noindent where $n_w^{(i)}$ is the number of water molecules
    around the $i^{th}$ particle and $\epsilon_w$ and $s_w$ are
    the interaction energies and entropy of the solvent,
    respectively.  The authors show\cite{kn:shiryayev_2005} that this solute-solvent
    interaction leads to an effective protein-protein square-well interaction
    whose strength is given by
    $\tilde{\epsilon}=\epsilon+2\epsilon_w-2k_BT\Delta s_w$.
    Thus,
    the Boltzmann weighting factor for the system with solvent is
    the same as the one without solvent, except that
    $\tilde{\epsilon}$ replaces $\epsilon$.  This yields several
    results for the model with solvent in terms of the behavior of
    the model without solvent.  For example, the phase diagram for
    the system with solute-solvent interactions is given by
      \begin{equation}
        k_BT_{coex} = \frac{(\epsilon_0 + 2 \epsilon_w)}{1+ 2\Delta s_w
        \tau(\rho)}\tau(\rho),
        \label{mapping}
    \end{equation}

    \noindent where $\tau(\rho)$ is the functional form of the
    coexistence curve (denoting two coexisting phases)
    without solvent. This
    expression is  independent of dimensionality.   Similarly, the radial distribution function
    for the system with
    solute-solvent interactions
    is given in terms of such interactions by
    \begin{equation}
        g_s(r) \equiv g_0(r;\tilde{\epsilon}).
    \end{equation}
    Correspondingly, the structure factor for the system with
    solute-solvent interactions is given by
     \begin{equation}
        S_s(q) = S_0(q;\tilde{\epsilon}).
     \end{equation}

     \noindent For a complete discussion of the model, the
    reader is referred to Ref. (\cite{kn:shiryayev_2005}).

    We obtain the fluid-fluid coexistence curves of $N = 250$ particles without
    solvent via Gibbs Ensemble\cite{kn:gibbs} Monte Carlo simulations.  We then use eq. (\ref{mapping}) to determine
    the coexistence curves
    of the particles interacting via eq. (\ref{square_well}) that
    includes the solute-solvent interactions.
    Equilibration runs
    lasted 10 millon Monte Carlo (MC) steps, production runs lasted 20
    million MC steps. Simulations were initialized from a liquid
    configuration.

    Isobaric-isothermal (NPT) Monte Carlo simulations\cite{kn:smit} were
    employed to sample the fluid and solid phases of the 2D system without
    solvent.  For
    sufficiently low ranges of $\lambda$, we observed a tendency
    for the fluid to crystallize and the solid phase to
    melt at appropriate pressures, as found in another study\cite{kn:noro}.
    To overcome this difficulty, we employ a biasing parameter which
    allows us to force the system into either a liquid or solid
    state. We use the bond-order parameter\cite{kn:steinhardt} defined by

    \begin{equation}
        Q_{lm} \equiv Y_{lm}(\phi(\vec{r})),
    \end{equation}

    \noindent where $Y_{lm}$ are the spherical harmonics. We
    calculate the bond order parameter \cite{kn:steinhardt}, $Q_6$, which has a non-zero
    value for a two dimensional solid and a zero value for the liquid. $Q_6$ is
    calculated~\cite{kn:steinhardt} as

    \begin{equation}
        Q_l = [\frac{4\pi}{2l +
        1}\sum_{m=-l}^{m=l}<Q_{lm}>^2]^{1/2}.
        \label{Q6}
    \end{equation}

    \noindent Less than 5\% of
    all configurations are rejected in any simulation of the
    liquid or solid. Using this order parameter we are able to sample fluid and solid
    phases and measure the radial distribution function, $g(r)$, and the bond-order correlation
    function, $g_6(r)$, given by

    \begin{equation}
    g_6(r) = <\psi_6^*(\vec{r}_j)\psi_6(\vec{r}_i)>
    \label{g6}
\end{equation}

\noindent where the local bond-order parameter $\psi(\vec{r}_i)$
for particle \emph{i} at position $\vec{r}_i$ is given by

\begin{equation}
    \psi_6(\vec{r}_i) = \frac{\sum_{k}
    w(r_{ik})\exp(i6\theta_{ik})}{\sum_{k}w(r_{ik})}.
    \label{psi6}
\end{equation}

 \noindent The summation is over neighboring particles \emph{k}
of particle \emph{i}; $\theta_{ik}$ is the angle between the
vector $(\vec{r}_i - \vec{r}_k)$ and a fixed reference axis. The
weighting factor \emph{w} is used to define nearest neighbors.

\section{Results and Discussion}
    The fluid-fluid coexistence curves have been calculated using
    Gibbs ensemble Monte Carlo.  Below $\lambda = 1.50$, we were unable to obtain a fluid-fluid coexistence
    curve. For particles interacting via a range of attraction of $\lambda = 1.375$, we
    observe that the expected fluid-branch of the coexistence
    curve
    tends to solidify, exhibiting a structure consistent with a solid. This was observed for the range $\lambda=1.25$
    as well. For $\lambda = 1.50$ and above, however, we do indeed observe fluid-fluid coexistence. This was verified by
    calculating both the radial distribution function, $g(r)$, and the bond-order correlation function, $g_6(r)$,
    calculated as described previously. Figs. \ref{figure1} and \ref{figure2} show this behavior for
    the various ranges considered. Similar behavior has been observed for
    another system
    of particles interacting via a short-range potential\cite{kn:noro} in 2D. In that study,
    this phenomenon
    was observed
    at short-ranges of attraction
    corresponding to the case where the spinodal curve is metastable with respect to the
    liquidus-solidus curve, suggesting that the binodal curve is metastable as well. It has
    been proposed \cite{kn:noro}
    that the free-energy barrier to be overcome is much smaller in 2D than in
    3D. Thus, at sufficiently low ranges of interaction, the
    metastable fluid-fluid coexistence curve cannot be sampled.

    The
    fluid-fluid coexistence curves for values $\lambda \geq 1.50$
    are shown in Fig. \ref{figure3}.
    It is worthwhile to compare these coexistence curves to their
    3D counterparts\cite{kn:vega}. The critical temperatures in 2D at these interaction ranges,
    $\lambda$,
    are relatively close to each other as compared to those in
    3D. Also, the critical temperatures are lower than those in
    3D.  The critical densities are different as well, yet they
    exhibit the same behavior as their 3D counterparts - the
    critical density increases with decreasing $\lambda$.

     To obtain the fluid-fluid coexistence curves with solvent we use eq.
    (\ref{mapping}). It has been shown\cite{kn:shiryayev_2005} that for appropriate
    choices of the parameters $\epsilon_w$ and $s_w$, a variety of
    phase diagrams can be realized. We choose the parameters $\epsilon_w =
    -1.0$ and $s_w = -1.50$ to see how the 2D fluid-fluid
    coexistence curves are affected in the presence of solvent
    with these particular interaction parameters. For these values of the parameters,
    it has been
    shown\cite{kn:shiryayev_2005} for the 3D square-well model that the
    phase diagram at $\lambda = 1.25$ has a lower critical point,
    whereas the model in the absence of solute-solvent interaction
    has an upper critical point. We find a similar behavior in 2D. It remains to be seen, however, if
    whether membrane proteins (or colloids) constrained to interact in quasi-two dimensional space exhibit
    such phase diagrams. Figure \ref{figure4} shows the fluid-fluid coexistence
    curves at the ranges we have studied, with
    solvent.

    As discussed previously, many agents are added to a
    protein-solvent mixture to control the protein-protein interaction range. The second virial
    coefficient
    coefficient, $B_2$,  is a measure of the net
    interaction and is commonly used as a means of determining the
 window for optimal globular protein crystallization \cite{kn:wilson}. This is presumably also true
 for membrane proteins. Therefore we show
    $B_2/B_2^{HS}$  as a
    function of temperature in Figure \ref{figure5}. ($B_2^{HS}$ is the second virial coefficient of
    hard spheres).  In the absence of solute-solvent interactions, $B_2 \rightarrow B_2^{HS}$ as
    $T \rightarrow \infty$, as the hard sphere interaction dominates.   However, in the presence
    of solute-solvent interactions  $B_2 \rightarrow
B_2^{HS}$ as
    $T \rightarrow \frac{1}{3}$.  This occurs because the effective
    interaction between particles, $\tilde{\epsilon}$, vanishes at that
    temperature, leaving only the hard sphere interaction.  (The
    same situation occurs in three dimensions.)  Although the second
    virial coefficient is only a function of temperature, if one
    plots its behavior along the coexistence curve, it becomes a
    function of density through the dependence of the coexistence
    temperature on density.  We show this dependence in Figure
    \ref{figure6}.  Along such a path the second virial coefficients
    with and without solute-solvent interaction are equal.

    The model studied here can be also used to study the
    case in which the effective interactions introduced by the
    solute-solvent interaction are
    temperature-dependent. This can be obtained from eq. (\ref{mapping}) by introducing
    temperature-dependent parameters
    $\epsilon_w$ and $\Delta s_w$. To illustrate this situation, we use the four-level model
    of Muller, Lee, and
    Graziano\cite{kn:muller,kn:lee,kn:moelbert} (MLG) for water, as it provides a simple example
    of
    such temperature dependent parameters. (We are not claiming that such a model is realistic in
    our case, but it does provide us with an illustrative example.) It can be shown\cite{kn:shiryayev_2005} that $\epsilon_w$ and
$\Delta s_w$
    can be obtained in the MLG model as
    $\epsilon_w(kT ) = E_s - E_b$ and $\Delta s_w(kT ) = S_s - S_b$ where the
    former and latter equations are the differences in energy and
    entropy in the bulk and shell states, respectively.  (See \cite{kn:moelbert} for
    the definition of these energies and entropies.) Fig. \ref{figure7} shows
    closed-loop fluid-fluid coexistence curves for choices of the
    parameter $\epsilon_0 = 0.5$ and $\epsilon_0 = 1.0$ in eq. \ref{mapping}.

    Finally, we briefly discuss the possibility of a hexatic phase for our model.
     To test for this
    possibility, we determine the radial distribution function $g(r)$ and
    the bond-order correlation function $g_6(r)$ for the fluid and solid phases at $\lambda =
    1.25$. Figures \ref{figure8} and \ref{figure9} show our results
    for typical values of pressure and temperature in the liquid
    and solid phases. Figure \ref{figure8} shows two typical g(r)
    plots corresponding to a fluid phase as well as one
    corresponding to a solid phase.  Figure
    \ref{figure9} shows our calculation of $g_6(r)$. For a
    hexatic phase, one would expect an algebraic decay. Our plot
    shows behavior typical of fluids and solids; $g_6(r)$ decays
    rapidly to zero in the fluid phase, indicating absence of long-range order, while that of
    the solid decays to a constant value.
    We see no
    evidence of a hexatic phase for our two dimensional model. However, because our system is quite
    small,
     finite
    size effects could be important.  Therefore our results are
    certainly  not definitive.

\section{Conclusion}

    We have studied the phase behavior of a simple model that is
    meant to describe the generic features of membrane proteins
    confined to a quasi-two dimensional geometry. In particular,
    we have shown that
   a simple phenomenological
    model for the solute-solvent interactions can yield phase diagrams with
    upper and  lower fluid-fluid critical points, as well as closed
    loops, depending on the choice of interaction parameters.

\section{Acknowledgements}  This work was supported by the G.
Harold Mathers and Leila Y. Mathers Foundation and by the National
Science Foundation, Grant DMR-0302598.

\newpage

\begin{figure}
    \center
     \rotatebox{-90}{\scalebox{0.5}{\includegraphics{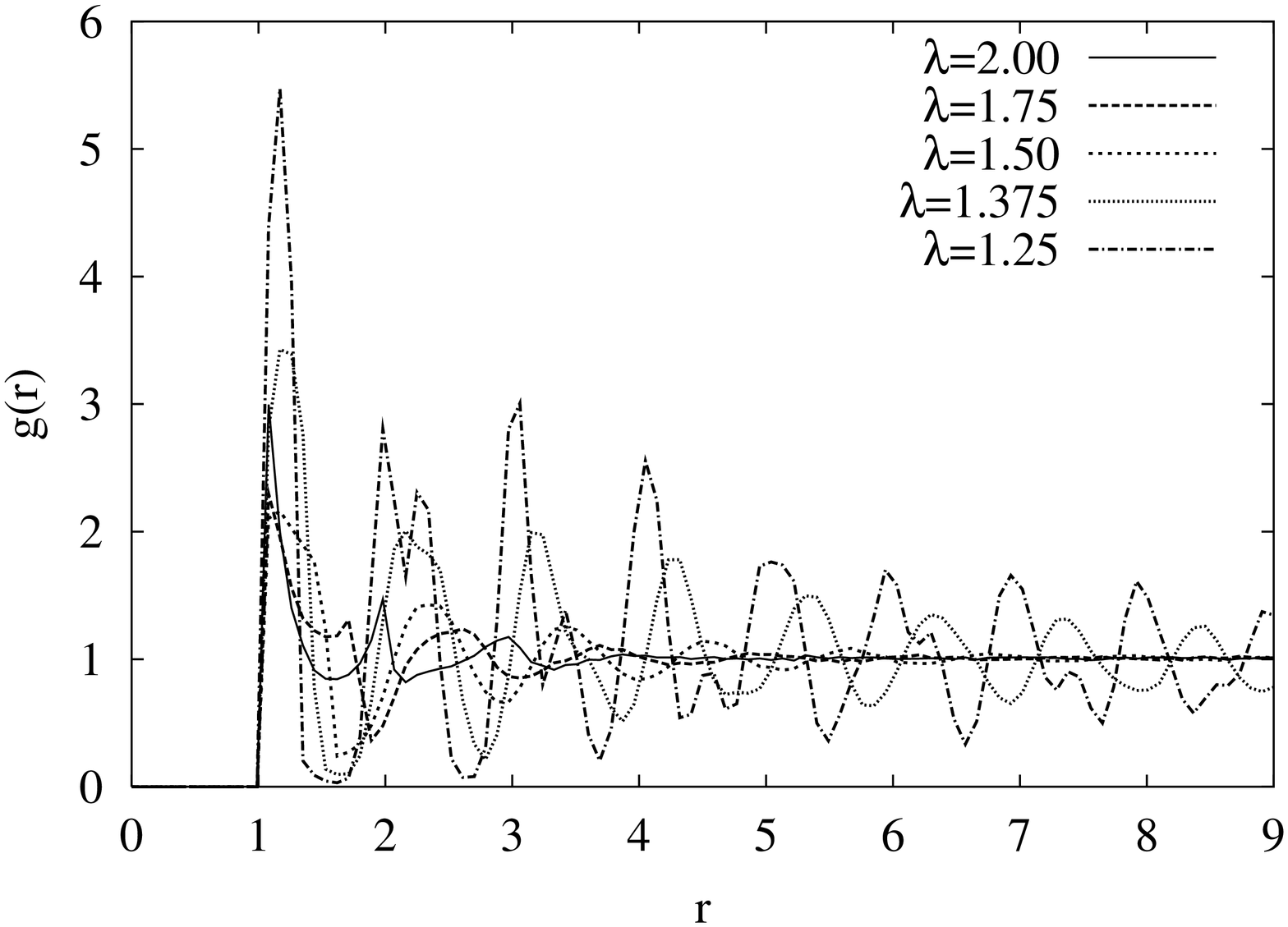}}}
     \caption{\label{fig:epsart}\small {}}
     \label{figure1}
    \end{figure}

\noindent Fig. \ref{figure1}: Radial distribution functions for
the ranges $1.25 \leq \lambda \leq 2.0$.
     Long-range order begins to appear for ranges where $\lambda \leq
     1.375$\\

\begin{figure}
    \center
     \rotatebox{-90}{\scalebox{0.5}{\includegraphics{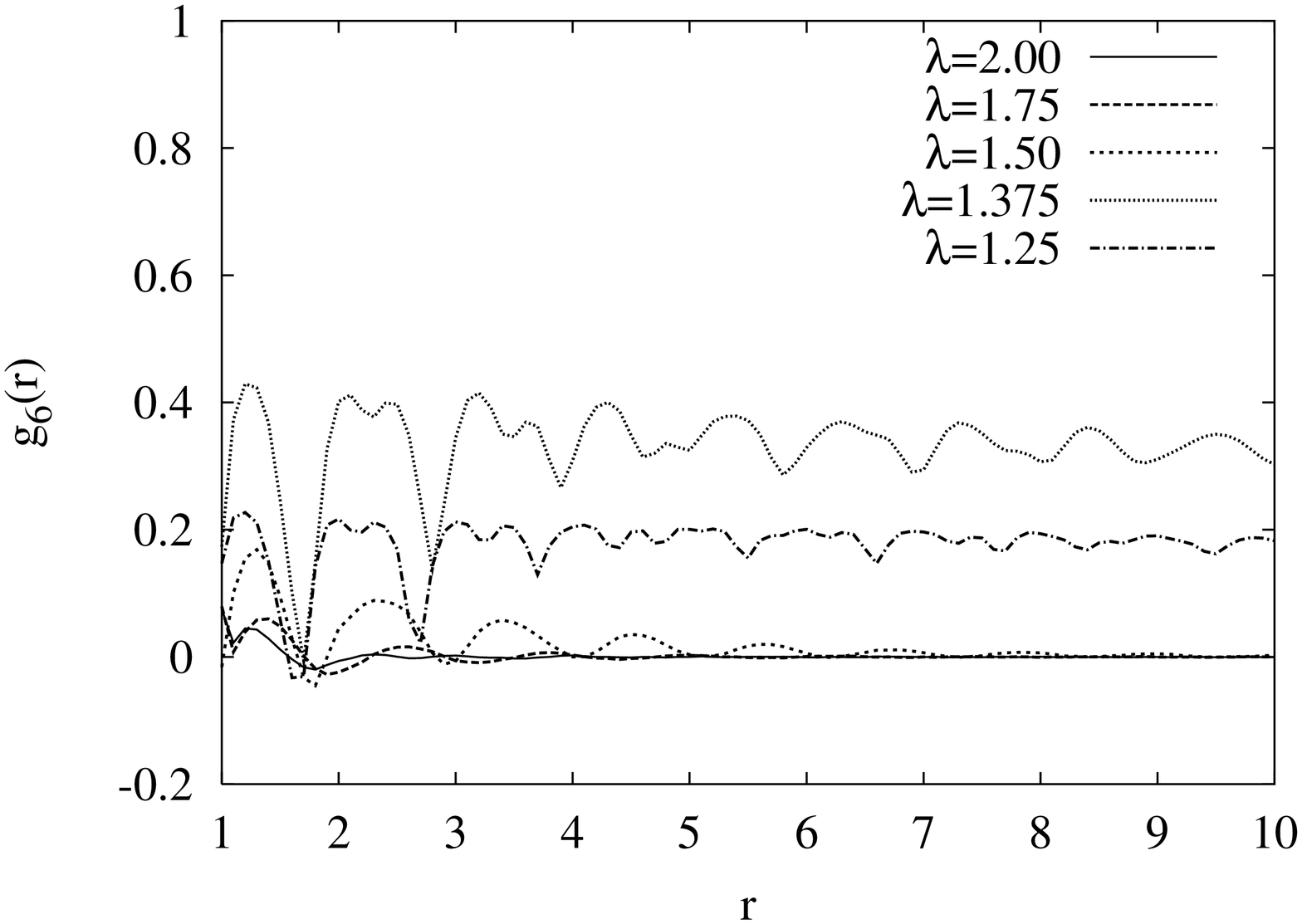}}}
     \caption{\label{fig:epsart}\small { }}
     \label{figure2}
    \end{figure}

\noindent Fig. \ref{figure2}: Bond-order correlation functions for
the ranges $1.25 \leq \lambda \leq 2.0$.
     A structure consistent with a solid is found for ranges where $\lambda \leq
     1.375$\\

\begin{figure}
    \center
     \rotatebox{-90}{\scalebox{0.5}{\includegraphics{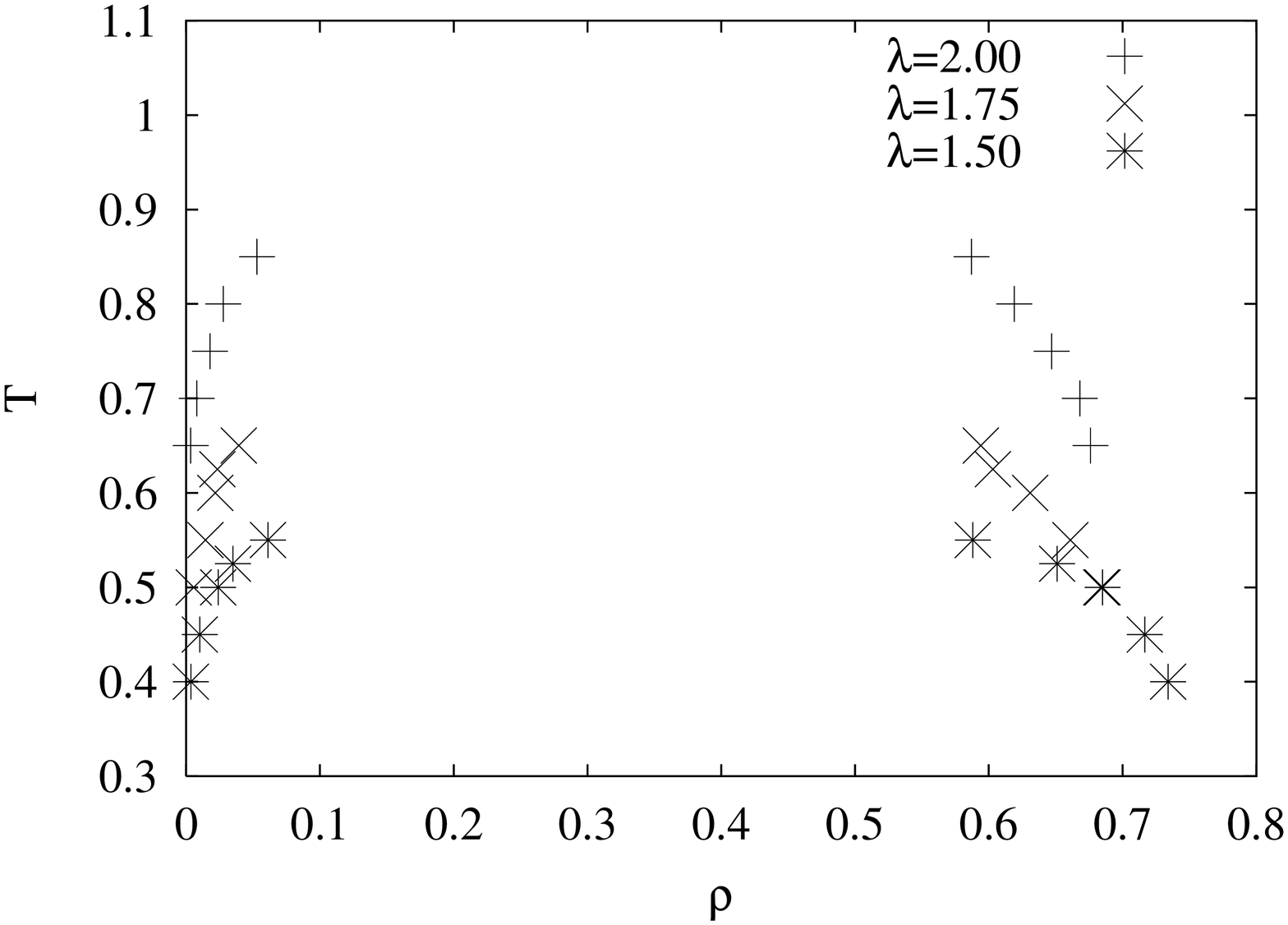}}}
     \caption{\label{fig:epsart}\small {}}
     \label{figure3}
\end{figure}

\noindent Fig. \ref{figure3}: Fluid-fluid coexistence curves for
the ranges
     $\lambda = 2.0$, $1.75$, and $1.50$ for the 2D
     square-well model.\\

\begin{figure}
    \center
     \rotatebox{-90}{\scalebox{0.5}{\includegraphics{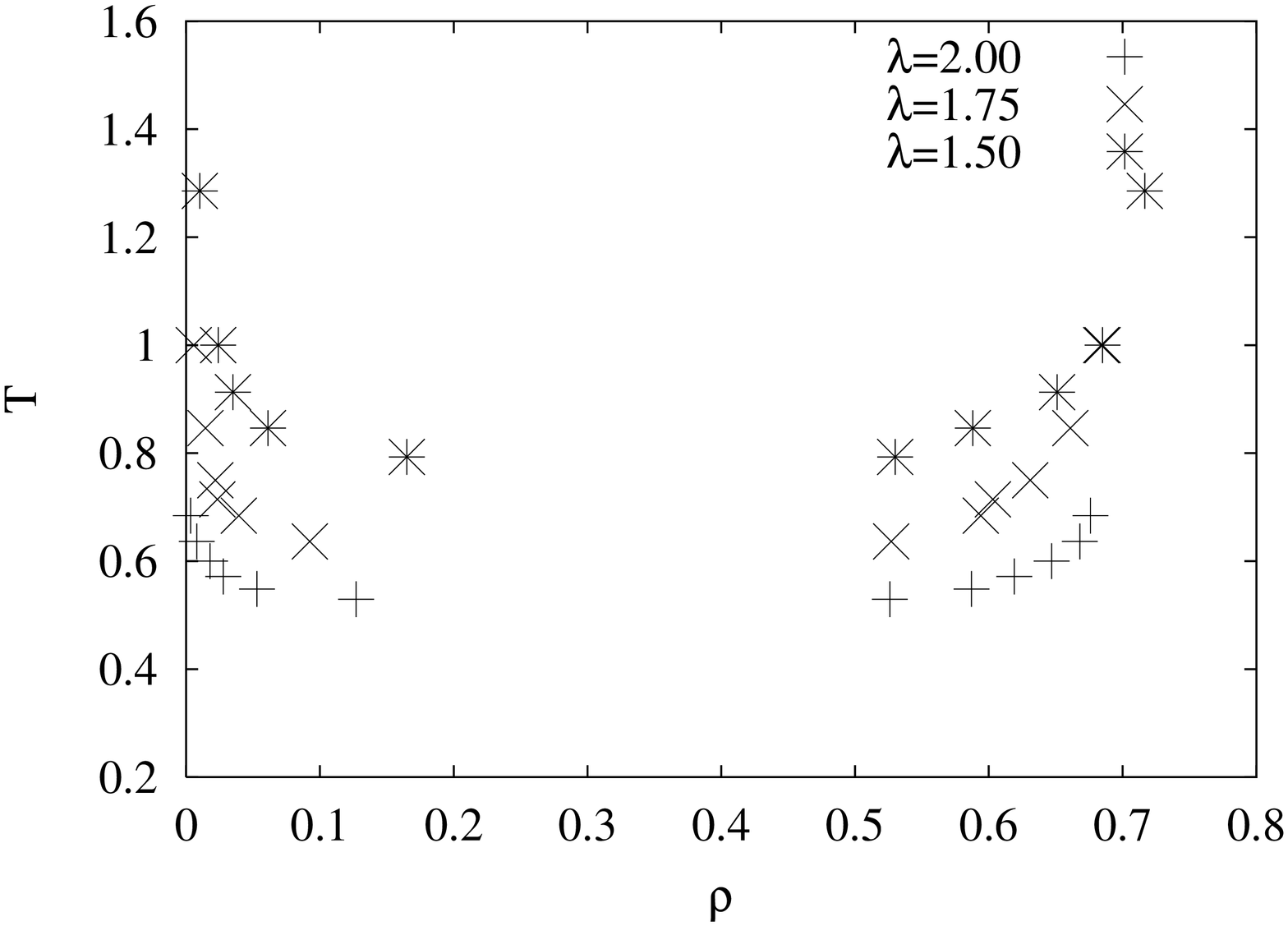}}}
     \caption{\label{fig:epsart}\small {}}
     \label{figure4}
    \end{figure}

\noindent Fig. \ref{figure4}: Fluid-fluid coexistence curves for
the
     ranges $\lambda = 2.0, 1.75$, and $1.50$ for the 2D
     square-well model in the presence of the solvent.\\

\begin{figure}
    \center
     \rotatebox{-90}{\scalebox{0.5}{\includegraphics{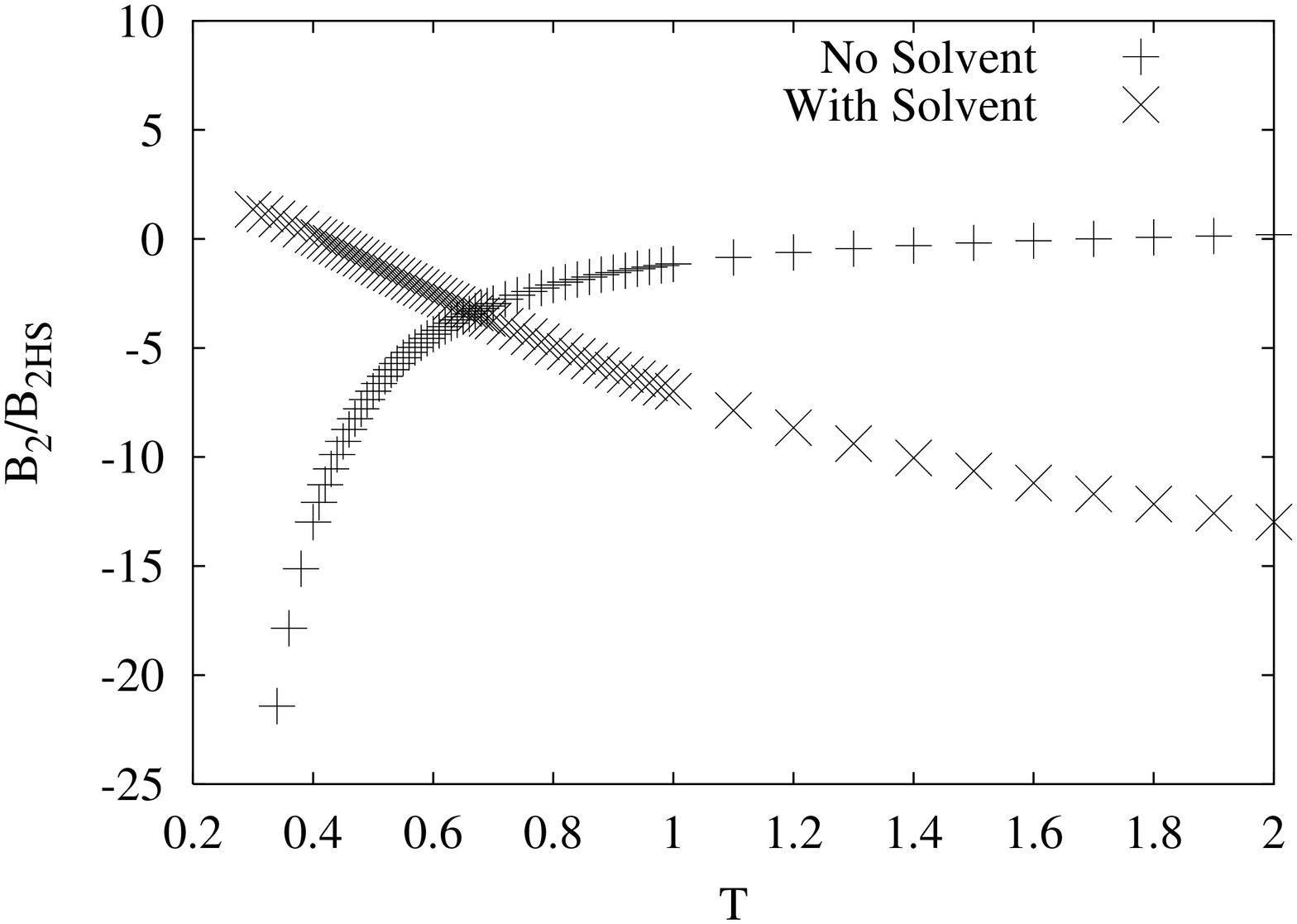}}}
     \caption{\label{fig:epsart}\small {}}
     \label{figure5}
    \end{figure}

\noindent Fig. \ref{figure5}: Second virial coefficient as a
function of
     temperature both in the absence and presence of the solvent.\\

\begin{figure}
    \center
     \rotatebox{-90}{\scalebox{0.5}{\includegraphics{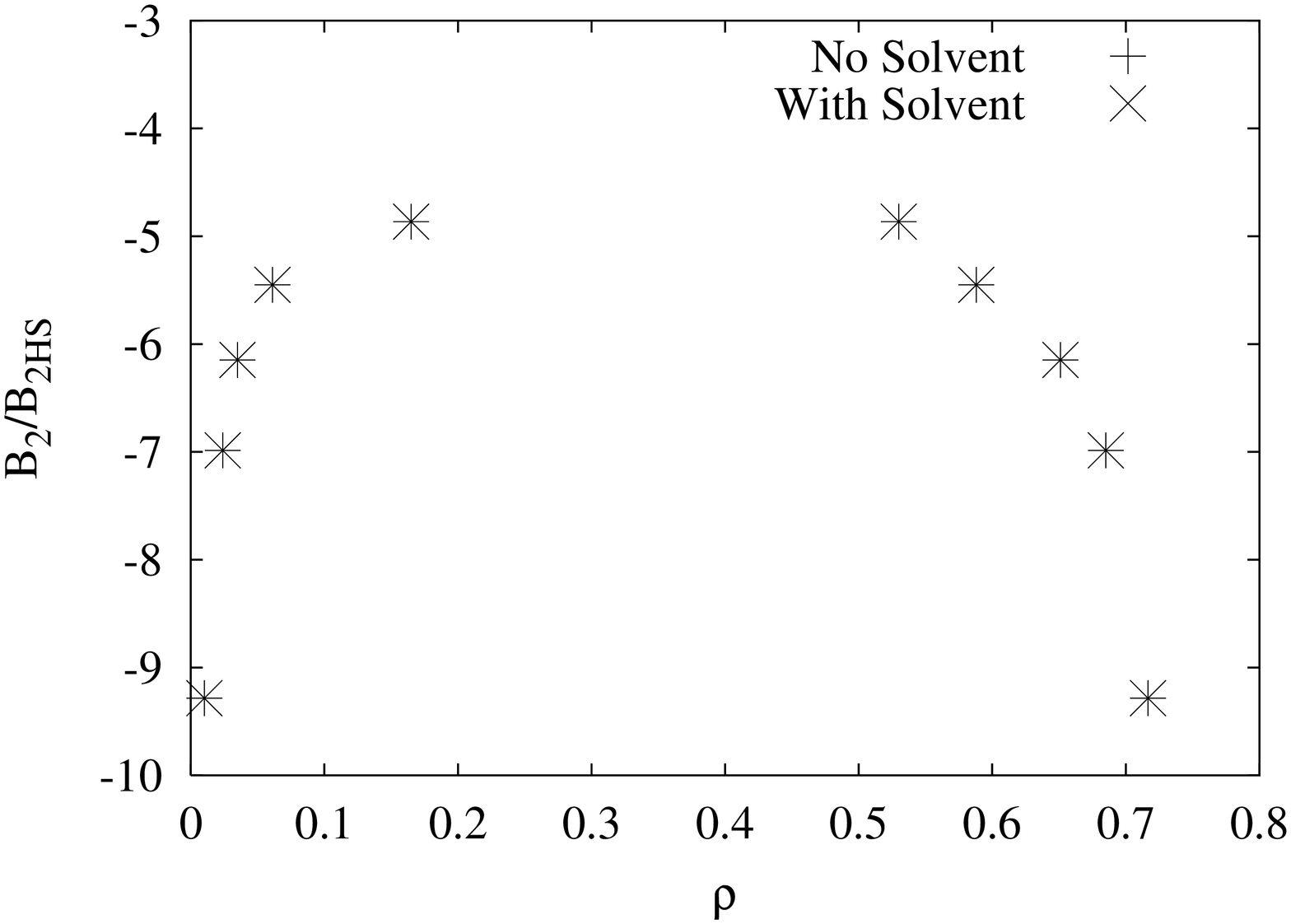}}}
     \caption{\label{fig:epsart}\small {}}
     \label{figure6}
    \end{figure}

\noindent Fig. \ref{figure6}: Second virial coefficient as a
function of density
      both in the absence and presence of the solvent.\\

\begin{figure}
    \center
     \rotatebox{-90}{\scalebox{0.5}{\includegraphics{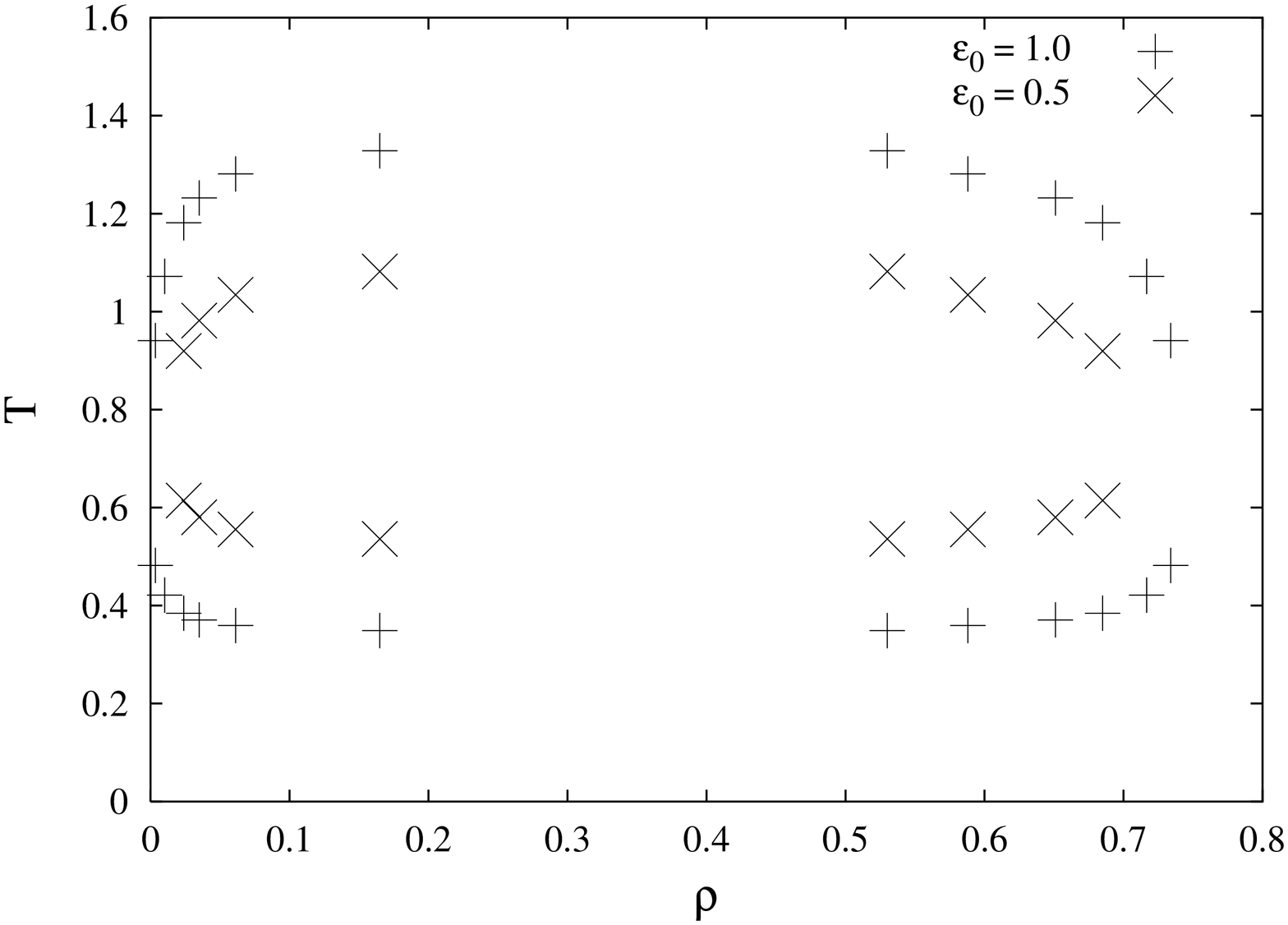}}}
     \caption{\label{fig:epsart}\small {}}
     \label{figure7}
\end{figure}

      \noindent Fig. \ref{figure7}: Fluid-fluid coexistence curves for the two-dimensional
     square-well model
     with solvent. The closed loop behavior is obtained using a simple model to illustrate
     the effects of temperature-dependent parameters in the model on the fluid-fluid coexistence curve, as discussed
     in the text.\\

\begin{figure}
     \center
     \rotatebox{-90}{\scalebox{0.5}{\includegraphics{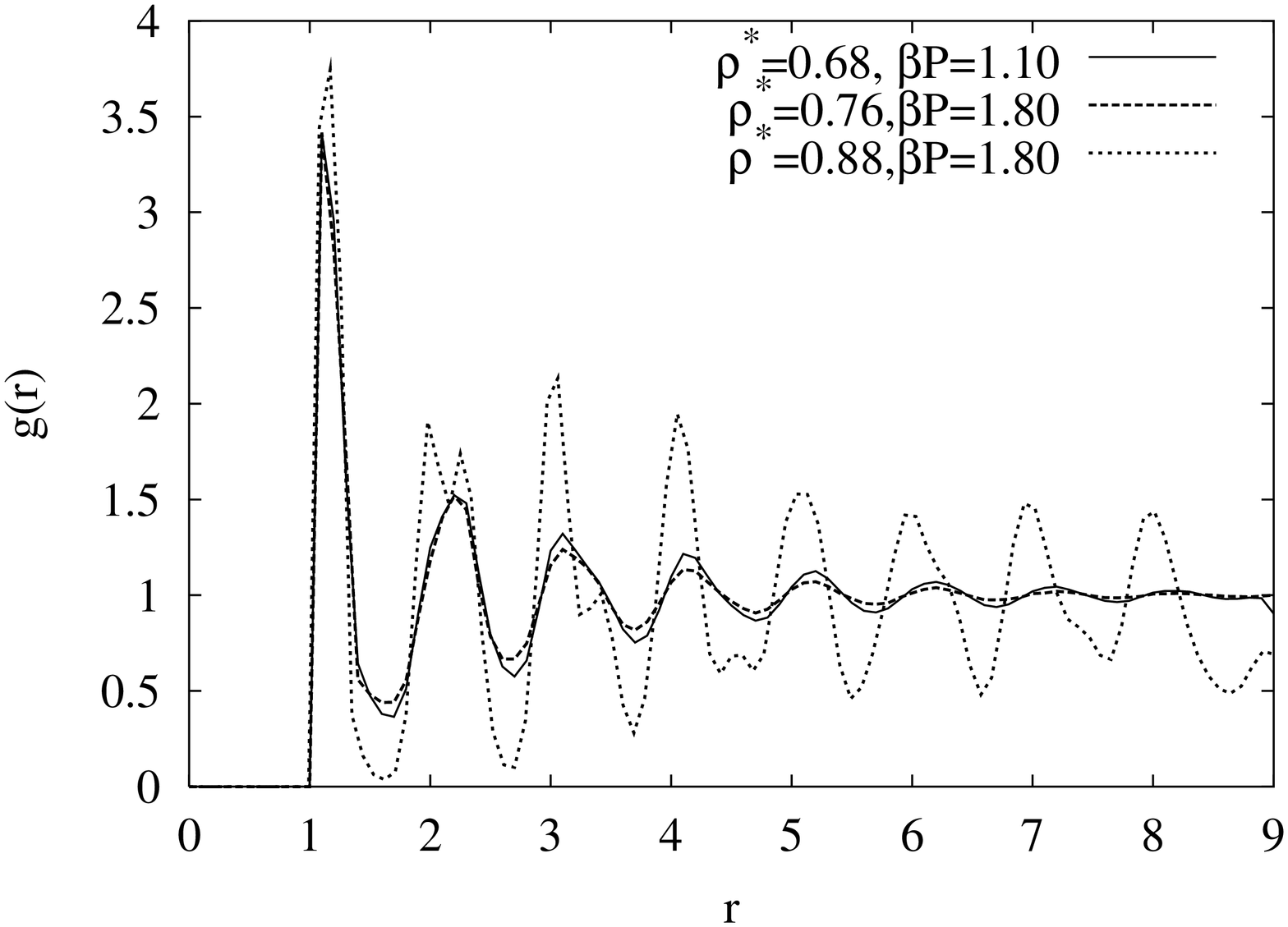}}}
     \caption{\label{fig:epsart}\small {}}
     \label{figure8}
    \end{figure}

      \noindent Fig. \ref{figure8}: Typical radial distribution functions for the liquid
     and solid phases at various densities $\rho$ and pressures
     $\beta$P.\\

\begin{figure}
     \center
     \rotatebox{-90}{\scalebox{0.5}{\includegraphics{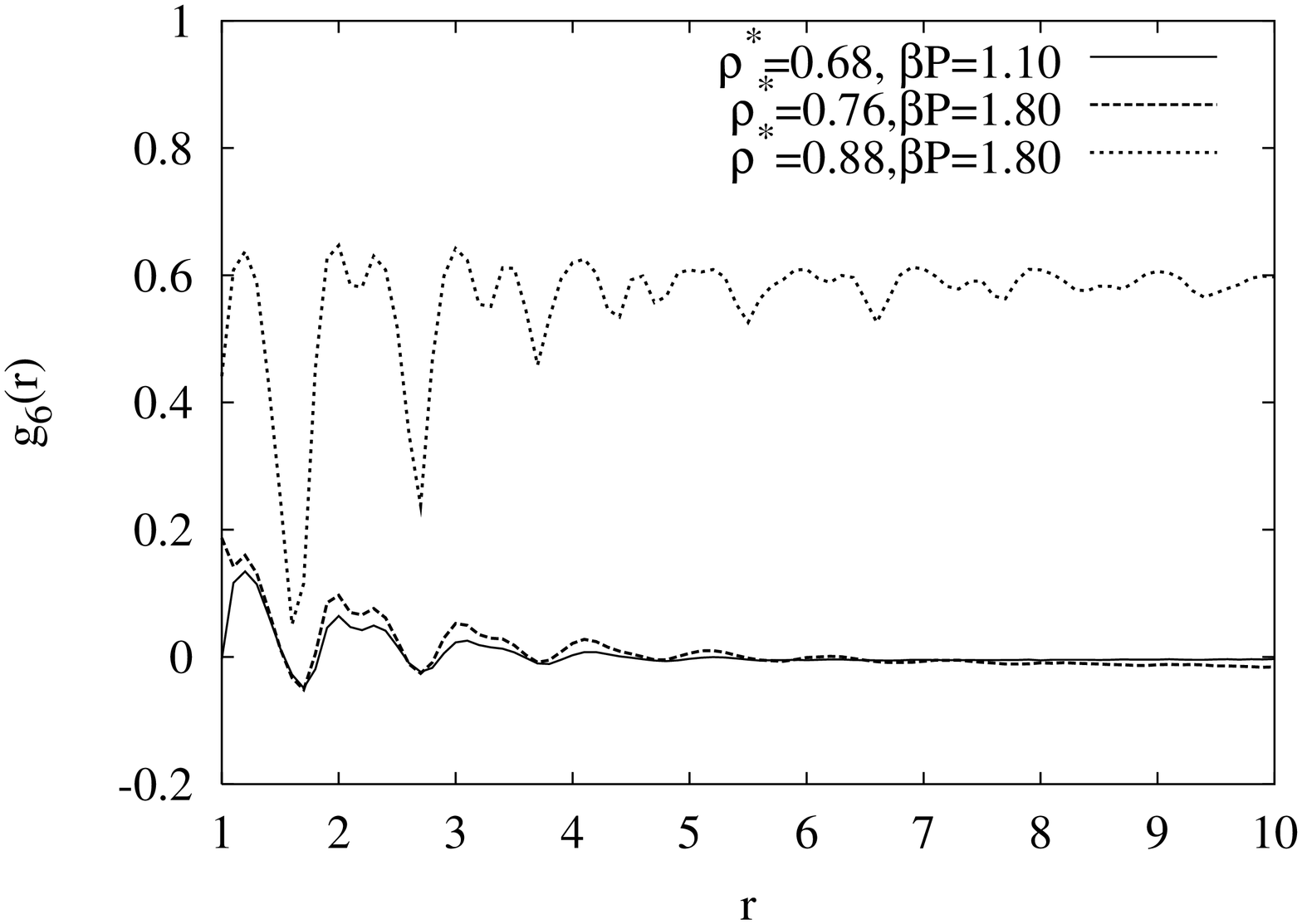}}}
     \caption{\label{fig:epsart}\small {}}
     \label{figure9}
    \end{figure}

      \noindent Fig. \ref{figure9}: Typical bond-order correlation functions for the
     liquid and solid
     phases at various densities $\rho$ and pressures $\beta$P.\\

\newpage

\end{document}